\documentclass[aps,prl,reprint,superscriptaddress,a4paper]{revtex4-1}

\usepackage[utf8]{inputenc}
\usepackage{amsmath}
\usepackage{amssymb}
\usepackage{graphicx}
\usepackage{hyperref}
\usepackage{color}
\usepackage{bm}

\begin{document}

\title{Quantum gas microscopy for single atom and spin detection}

\author{Christian Gross}
\affiliation{Max-Planck-Institut f\"{u}r Quantenoptik, 85748 Garching, Germany}
\affiliation{Munich  Center  for  Quantum  Science and Technology (MCQST), 80799 Munich, Germany}
\affiliation{Physikalisches Institut, Eberhard Karls Universit\"{a}t  T\"{u}bingen, 72076 T\"{u}bingen, Germany}
\author{Waseem S. Bakr}
\affiliation{Department of Physics, Princeton University, Princeton, New Jersey 08544, United States}

\date{\today}

\begin{abstract}
A particular strength of ultracold quantum gases are the versatile detection methods available. Since they are based on atom-light interactions, the whole quantum optics toolbox can be used to tailor the detection process to the specific scientific question to be explored in the experiment. Common methods include time-of-flight measurements to access the momentum distribution of the gas, the use of cavities to monitor global properties of the quantum gas with minimal disturbance and phase-contrast or high-intensity absorption imaging to obtain local real space information in high-density settings. Even the ultimate limit of detecting each and every atom locally has been realized in two-dimensions using so-called \textit{quantum gas microscopes}. In fact, these microscopes not only revolutionized the detection, but also the control of lattice gases. Here we provide a short overview of this technique, highlighting new observables as well as key experiments that have been enabled by quantum gas microscopy. 
\end{abstract}

\maketitle

\section*{Introduction}

Precise detection techniques are central to modern experimental physics. On the one hand they are key to testing physical theories with ever increasing precision, on the other hand they provide detailed insights into complex physical processes. In the context of quantum many-body physics one may argue that the ultimate detection limit is given by the detection of each and every constituent of the system. Taking the example of strongly correlated materials, this would translate into the detection of each electron in the material. Quantum simulators for such materials and for strongly correlated quantum many-body physics in general can be realized with ensembles of ultracold atoms~\cite{bloch2012,georgescu2014,gross2017}. Effective interaction control is achieved by direct control of their two-body interactions using Feshbach resonances~\cite{chin2010}, by suppressing the kinetic energy relative to the interaction energy as done in optical lattices, or by a combination of both~\cite{bloch2008a}. This allows for the realization of designer many-body systems, in which fundamental effects can be studied in a fully controlled, pristine environment. While detecting each electron inside a material seems like science-fiction, this ultimate detection limit can be reached in atomic quantum simulators through the local detection of each and every atom.

Traditional detection methods for ultracold gases allow for the extraction of the real or momentum space distribution by resonant or near-resonant laser beam transmission or by bulk fluorescence detection~\cite{ketterle1999a,ketterle2008}. These methods are powerful and relatively simple to implement with fairly high spatial resolution~\cite{gemelke2009,cocchi2016}, yet they do not reach local single atom sensitivity in high-density settings. For individual atoms or in sparse samples, however, high-fidelity single atom detection has been implemented already at the beginning of the 21st century~\cite{schlosser2001,miroshnychenko2003,nelson2007} and atom counting has been demonstrated~\cite{serwane2011,hume2013}. An extreme resolution, albeit with low single atom detection fidelity, can be achieved using focused electron beams~\cite{wuertz2009}. Experiments using metastable helium condensates achieve similar efficiencies for single atom detection ($10\%$ to $25\%$) in momentum space imaging using electron multipliers and microchannel plates~\cite{lopes2015,hodgman2017}. With the invention of so-called \textit{quantum gas microscopes} in 2009, single atom detection with near unity detection efficiency became available in a many-body environment~\cite{bakr2009,sherson2010}. Similar to standard fluorescence detection, quantum gas microscopes detect the individual atoms optically by observing scattered photons.  For half a decade only  microscopes for bosonic rubidium were available due to the more demanding imaging for the fermionic alkali isotopes potassium and lithium. In 2015, however, several experiments reported the imaging of individual fermions in a lattice~\cite{cheuk2015, haller2015, parsons2015, edge2015} and soon after the local observation of Pauli blocking~\cite{omran2015}, Mott insulators~\cite{greif2016, cheuk2016b} and antiferromagnetic spin correlations in Hubbard systems~\cite{cheuk2016a, parsons2016,boll2016, brown2017} has been reported, which have been previously measured without single atom sensitivity~\cite{truscott2001,sanner2010,mueller2010,joerdens2008, schneider2008,greif2013,hart2015}.  The microscopes offer access to observables, which were previously inaccessible in any itinerant quantum many-body system. Most prominent are the full counting statistics and non-local multi-point correlators, which have enabled the direct detection of string order~\cite{endres2011}, entanglement entropy~\cite{islam2015,kaufman2016,lukin2019,rispoli2019}, hidden correlations~\cite{hilker2017} and magnetic polarons~\cite{koepsell2019,chiu2019}. Techniques to measure similar observables in momentum space have also been developed~\cite{gring2012, schweigler2017}.

In ultracold atoms setups, high resolution optical detection also enables high resolution control. In fact, single atom and spin manipulation is now possible even in samples where the constituents are spaced only half a micrometer apart~\cite{weitenberg2011b,zupancic2016}. This technique opens new routes in ultracold atom research, in particular for quantum simulation applications, where local potential programming is now available, or for quantum information, where local spin manipulation is fundamental to the implementation of single qubit gates~\cite{weitenberg2011c}.

In this short review we will discuss these new high resolution detection and manipulation techniques with a focus on short-spacing optical lattice systems. For a complete picture we point the reader towards the related reviews on optical tweezers [citation to be added], trapping and control in bulk samples [citation to be added] and new time-of-flight and spectroscopical detection methods [citation to be added].

\section*{Optical lattices}

Optical lattices are central to the quantum gas microscope technique. They serve two conceptually different purposes: Less fundamental, it is often the physics in a lattice that is explored in quantum gas microscopes, which is motivated by the fact that atoms are localized to discrete positions and atom-by-atom detection translates into the counting of atoms at each of the lattice sites. More fundamental is the need for lattices during detection. Here, they pin the atoms in place during fluorescence detection, such that enough photons can be scattered per atom to reach a sufficiently high single atom signal-to-noise ratio. Optical lattices and the many-body physics they enable have been reviewed in detail, for example, in~\cite{bloch2008a} and we limit the discussion here to what is needed to understand single atom detection in them.\\

Short-spacing optical lattices are formed by interfering laser beams of wavelength $\lambda$ under a large angle. This results in an interference pattern with minimal distance $d=\lambda/2$ for head-on interference. By superposition of such interfering beam pairs from several directions one can form a lattice of intensity maxima, in which the atoms can be tightly confined. Given the large angles involved, this distance $d$ is typically slightly below the optical resolution limit at the typical wavelengths of suitable atomic dipole transitions. In order to hold the atoms in place while scattered photons are collected, the lattice depth $V_0$ has to be much larger than the temperature of the atoms during imaging. This is orders of magnitude higher than the temperature, at which ultracold physics takes place and, thus, the quantum ensemble is destroyed in the detection process. Typically, deep lattices with $V_0$ on the order of hundreds of microkelvins are required, which is an important consideration for experiment design. Methods to reach such large optical trap depths include the use of high power lasers with tens of Watts output power~\cite{sherson2010,cheuk2015,haller2015,parsons2015,omran2015,miranda2015,edge2015,yamamoto2016,brown2017} or dedicated near-resonant lattices used just for imaging~\cite{bakr2009}.  Another important characteristic for several imaging schemes is the size of the onsite ground state wave function as measured by its Gaussian standard deviation $\Delta x = \sqrt{\hbar/m \omega}$, with atomic mass $m$ and onsite trapping frequency $\omega$. The onsite trapping frequency $\hbar \omega= 2 E_r \sqrt{V_0/E_r}$ is determined by the characteristic energy scale $E_r=h^2/8 m d^2$ associated with the lattice. The onsite size $\Delta x$ compared to the imaging wavelength determines the performance of Raman-sideband cooling via the Lamb-Dicke parameter~\cite{monroe1995a,hamann1998,kerman2000} or the need to phase-modulate the optical lattice in an optical molasses scheme, for which the thermal scale of the wave packet size is $\sqrt{\frac{k_B T}{m \omega^2}}$~\cite{sherson2010}.

Naively one might expect the fact that optical lattice spacings are typically sub-wavelength scale makes onsite detection of individual atoms impossible. The solution to this challenge are image reconstruction algorithms, which use the discrete spacing of the atoms as prior knowledge. Given that enough photons can be scattered from individual atoms, the algorithm can reconstruct the absolute position of the lattice on the image and analyze fluorescence levels locally. For typically reached signal to noise levels this results in a fidelity well above $99\%$ to distinguish between zero and one atom on a site.

\begin{figure*}
    \centering
    \includegraphics[width=1\textwidth]{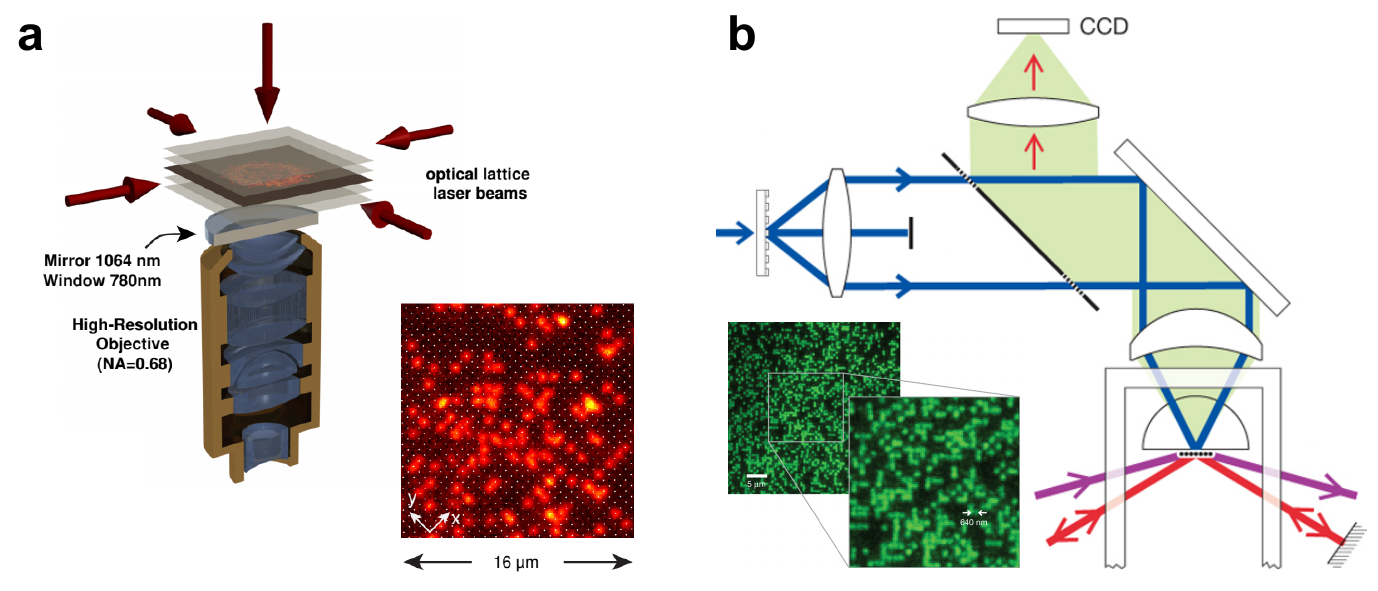}
    \caption{\textbf{Quantum gas microscopy.} The two most prominent schemes to achieve high numerical aperture in an ultra-high vacuum environment are illustrated here. \textbf{a} A custom objective outside the vacuum chamber allows one to work far away from surfaces. The image shows an exemplary fluorescence image in which individual Rubidium atoms in the optical lattice are clearly distinguishable as colored ``blobs''. \textbf{b} A superpolished half-sphere inside the vacuum chamber provides high NA close to its surface. The holographically generated optical lattice is projected ``backwards'' through the imaging system. Panel a has been originally published in~\cite{sherson2010} and panel b in~\cite{bakr2009}.}
    \label{fig1}
\end{figure*}

\section*{Quantum gas microscopy}

The high resolution imaging optics used in quantum gas microscopes needs to be compatible with the ultrahigh vacuum environment in which the atoms are held. This limits the practically usable numerical aperture. Two approaches for high resolution imaging are used: In the first, the atoms are brought within a distance of about ten microns to a superpolished half-sphere, which together with correction lenses downstream the optical path can result in a very high numerical aperture above 0.8~\cite{bakr2009,cheuk2015,parsons2015,miranda2015}. In the second approach, the atoms kept far away from any surface, but a customized objective is required to reach high numerical apertures (between 0.5 and 0.8)~\cite{sherson2010,haller2015,omran2015,edge2015,yamamoto2016,brown2017}. While optical access for the manipulation of the gas is better in the second case, the first can utilize the available laser power for optical lattice formation more efficiently, which simplifies the detection. The different schemes are illustrated in figure~\ref{fig1}.

With the need for high numerical apertures, the resulting depth of focus of the detection system of about a micron is small. Hence, only a single plane can be fully in focus of a single imaging path. To avoid deterioration of the signal, scattering from nearby planes has to be avoided, either by working with a single plane only~\cite{bakr2009, sherson2010,cheuk2015,haller2015,parsons2015,omran2015,miranda2015,edge2015,yamamoto2016,brown2017}, or by transporting additional planes far away from the plane of interest before imaging~\cite{koepsell2020}. Preparing an ultracold quantum gas in a single plane is an experimentally quite demanding task. It is achieved either by optical compression of the gas and subsequent loading into a lattice with potentially variable spacing~\cite{parsons2015,omran2015,miranda2015,brown2017}, evanescent wave traps~\cite{bakr2009} or by radio-frequency assisted removal of atoms in all planes but one~\cite{sherson2010,haller2015,edge2015,cheuk2015,yamamoto2016}. Often evaporative cooling after preparation of the single plane is required to reach the coldest temperatures of a few nanokelvins.

Central to sub-wavelength lattice resolution is the need to scatter many  photons per atom ($10^3$ to $10^5$) while keeping the atom at its place. As photon scattering heats the atom, this requires finding schemes that result in high photon scattering rates, but at the same time limit the temperature by cooling. The methods include bright molasses cooling used for rubidium~\cite{bakr2009, sherson2010}, Raman-sideband~\cite{cheuk2015,parsons2015,omran2015,brown2017} or EIT~\cite{haller2015,edge2015} cooling used for lithium or potassium and narrow-line cooling for ytterbium~\cite{yamamoto2016}. The details of the implementation of these methods are species dependent and we refer the reader to the original publications. All of these techniques necessarily involve near resonant light. This light induces molecule formation by photoassociation if two or more atoms are present on a single lattice site. This prevents imaging of more than two atoms per site with acceptable fidelity. In order to still access onsite observables with high fidelity, experiments use optimized photoassociation as a tool to ensure that all pairs of atoms per site get converted to molecules and are therefore not imaged. Quantum gas microscopes, thus, detect the parity of the atom number on each site of the optical lattice~\cite{bakr2009,sherson2010,endres2013}.

Quantum gas microscopes so far are mostly used to study the physics in optical lattices. However, the ``physics lattice'' does not necessarily need to be the same as the ``pinning lattice'' used during imaging~\cite{bakr2009}. In particular, the latter can be much shorter spaced, even if its sites cannot be optically resolved~\cite{shotter2011,omran2015}. This separation of physics and detection lattices has several advantages. First, the parity detection limit can be overcome by supersampling the physics lattice sites. Second, more complex physics lattice geometries can be realized, which are not compatible with the high power needed for imaging. One important example are optical superlattices~\cite{sebby-strabley2006,foelling2007}. These consist of two superposed lattices with precisely a factor of two different lattice spacings. The resulting lattice has a unit cell of two sites, whose geometry can be precisely controlled via the relative phase and power of the two fundamental lattices. The possibility to dynamically control the local geometry can be used as a powerful detector. An important example, demonstrated so far only without local resolution, is the detection of the singlet or triplet fraction~\cite{trotzky2010}. But also for quantum gas microscopy the superlattice is a powerful tool. In combination with magnetic fields it can be used to implement detection methods, which measure the local atom number and spin state in a single experimental run~\cite{boll2016}. This provides access to novel observables, foremost, to local and non-local correlations involving the density and the spin~\cite{hilker2017,salomon2019,koepsell2019}. For example, this opens a new window into the physics of the Hubbard model, where the complex interplay of spin and density is at the heart of the puzzle of unconventional metals and high-temperature superconductors~\cite{lee2006}.

Though cooling of the atoms is necessary in most current quantum gas microscopy schemes, it is not a fundamental requirement. If the atom of choice allows it, one can utilize short wavelengths for detection, or -- given the excellent control available for optical lattices -- the spacing of the atoms could be enlarged dynamically to separate the atoms before imaging. With non-overlapping point spread functions, few photons are sufficient to detect individual atoms~\cite{miranda2015,yamamoto2016,miranda2017}. This has been already demonstrated in the context of correlation detection in momentum space, where individual atom signals have been reconstructed~\cite{buecker2011,bergschneider2019}.

One important aspect of quantum gas microscopes is the ability to count individual excitations and quantum fluctuations of an atomic Mott insulator. This provides a precision thermometer and a precision detector for out-of-equilibrium physics, which has been used to study the spreading of quantum correlations after quenches~\cite{cheneau2012, fukuhara2015}, amplitude modes of the strongly interacting superfluid~\cite{endres2012a}, photon-assisted tunneling~\cite{ma2011} and universal prethermalization in driven systems~\cite{rubio-abadal2020}.

\section*{Local control}

\begin{figure*}
    \centering
    \includegraphics[width=\textwidth]{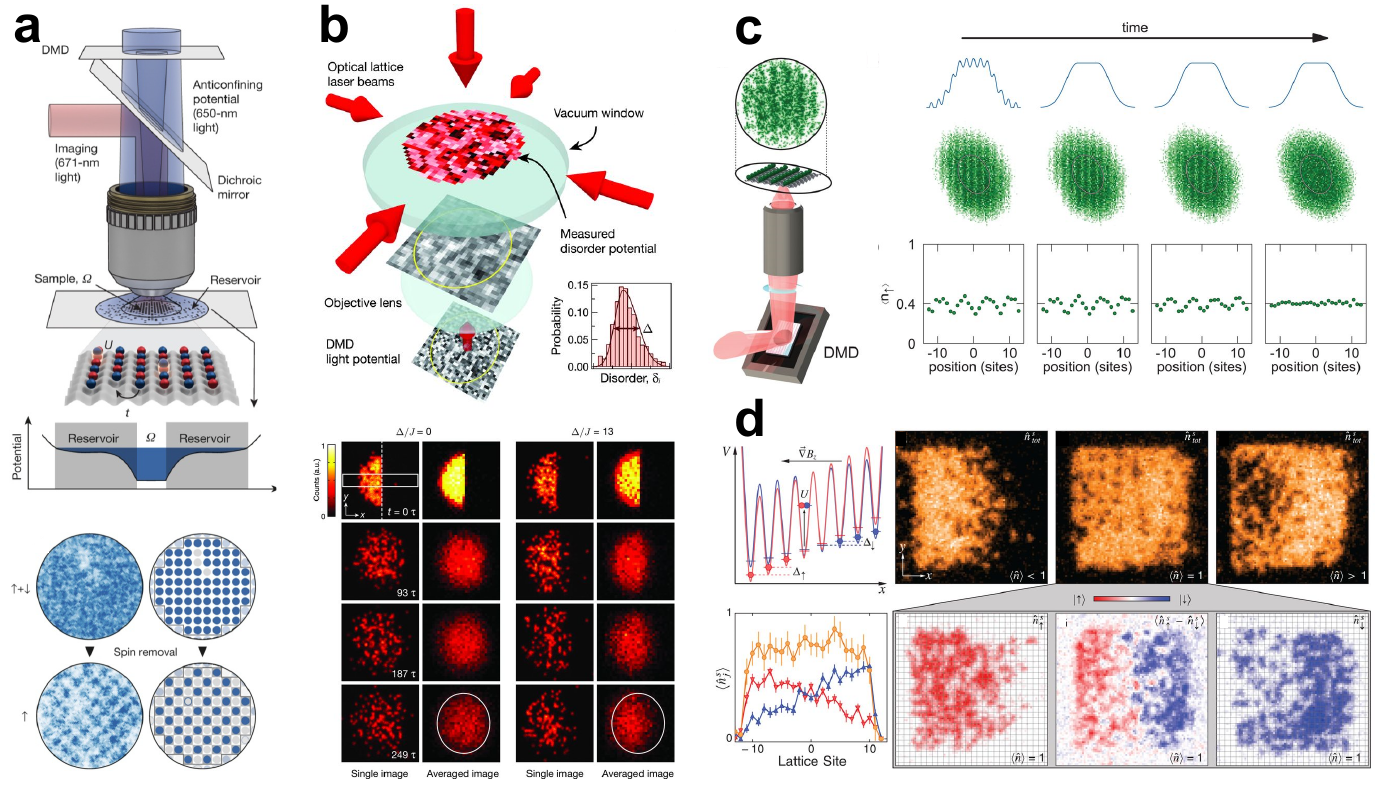}
    \caption{\textbf{Equilibrium physics and dynamics of strongly interacting lattice fermions.} \textbf{a} Hubbard model in equilibrium: Low enough temperatures for long-range antiferromagnetism have been reached by entropy engineering of a 2D lattice cloud. For vanishing doping, magnetic correlations over the entire system of about 100 sites have been observed.  \textbf{b} Dynamics of disordered systems: Many-body localization, the absence of thermalization in quantum many-body systems, has been studied in a disordered lattice. The disorder landscape can be precisely characterized by  spectroscopy. Starting from an engineered far-from equilibrium density distribution, localization manifests itself in an imbalanced steady state density (bottom right picture pair), while the distribution is symmetric in a clean system (bottom left picture pair). \textbf{c} Transport in the charge sector: Using a spatial light modulator, initial density densities with varying modulation length scales are prepared and their subsequent decay is monitored. Non-fermi liquid transport properties have been observed in this way in two-dimensional doped Hubbard systems. \textbf{d} Transport in the spin sector: By preparation of an initial spin imbalance in a uniform system, spin transport in a Hubbard layer has been studied. The spin diffusion coefficient can be precisely measured providing a benchmark for different theories. Panel a has been originally published in~\cite{mazurenko2017}, panel b in~\cite{choi2016}, panel c in~\cite{brown2019} and panel d in~\cite{nichols2019}. }
    \label{fig2}
\end{figure*}

The high resolution objective required for detection can also be used ``backwards'' to focus laser beams to sub-micron spot sizes. In particular, when studying lattice physics this provides a powerful tool to optically manipulate the atoms with the ultimate resolution of a single lattice site~\cite{weitenberg2011b,zupancic2016}. Different schemes based on spatial light modulators can be implemented to generate programmable local potentials.

Early quantum gas microscope experiments created engineered potential landscapes for atoms by ``projecting" light patterned with static lithographic masks. For example, this was used for for realizing a far-off-resonant physics lattice spatially registered with a much deeper near-resonant pinning lattice by illuminating the same mask with light at different wavelengths~\cite{bakr2009}. Later experiments utilized digital micromirror devices to create potentials that can reprogrammed on the fly. While each pixel of such a device can only be either on or off, grayscale control of the light intensity can be achieved by mapping tens or hundreds of pixels to the same site. Potential engineering with such devices allowed the optimization of entropy redistribution schemes in Fermi-Hubbard systems to achieve long-range antiferromagnets~\cite{mazurenko2017}. The related capability of engineering the initial density of the gas has been particularly useful for transport experiments, e.g. for studying many-body localization in disordered lattices~\cite{choi2016, rubio-abadal2019}, strange metallicity in clean Fermi-Hubbard models~\cite{brown2019} and non-diffusive transport in tilted Hubbard~\cite{guardado-sanchez2020} and Heisenberg spin systems~\cite{hild2014}. Further control of the projected light field was achieved by using digital micromirror devices in a Fourier rather than an image plane~\cite{islam2015}. This enables shaping both the phase and amplitude of the electric field, which allows for the correction of microscope aberrations. This mode also uses the available light power more efficiently when it is desirable to address only a small region of the cloud.

When the lattice gas consists of atoms prepared in more than one hyperfine state, the high resolution of a microscope may be used for rotating individual spins in the system. In the heavier alkali atoms, this is accomplished by using light with a frequency detuning that is small compared to the fine structure splitting, leading to significant differential Stark shifts for atoms in different spin states~\cite{rubio-abadal2019}. If such light is focused on a site, the spin state on that site can be rotated using a global microwave pulse that resolves the frequency shift of the transition. This has been used for example to study the quantum dynamics of free and bound magnon excitations in Heisenberg spin chains~\cite{fukuhara2013a, fukuhara2013} or to realize quenches by local removal of one of the spin states to trigger dynamics of few particles~\cite{preiss2015a,tai2017} or in a many-body environment~\cite{vijayan2020a, ji2020}. Such schemes addressing individual sites of a lattice are particularly challenging to implement in setups where the optical lattice is created using the traditional approach with retroreflected laser beams rather than projection through the microscope. While the former approach produces cleaner lattice potentials, the lattice can drift relative to the addressing beam. Nevertheless, feedback techniques have been used to track the drift of the lattice phase between experimental shots, allowing an addressing resolution of about a tenth of a lattice site~\cite{weitenberg2011b}.

\section*{New directions}

\begin{figure*}
    \centering
    \includegraphics[width=\textwidth]{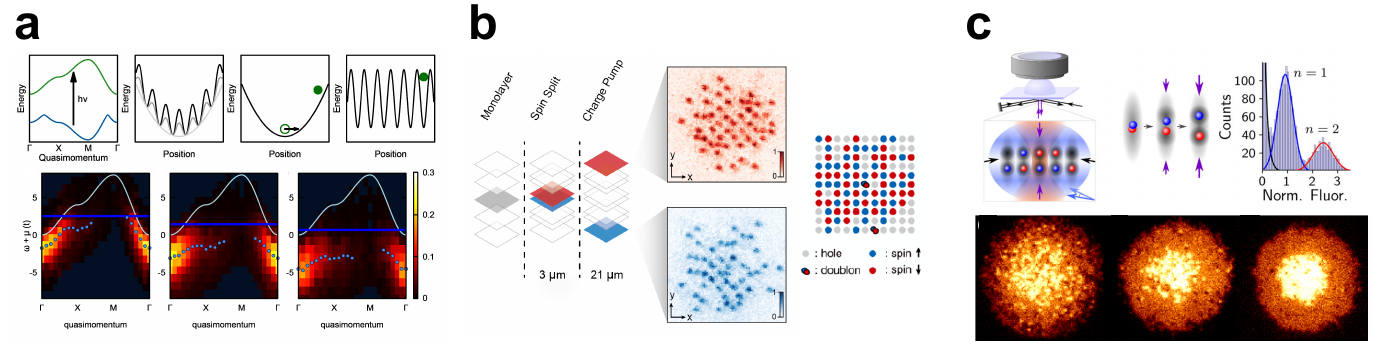}
    \caption{\textbf{New directions for quantum gas microscopy.} \textbf{a} High resolution ARPES is used to map out the momentum-resolved single-particle excitation spectrum across the Brillouin zone in an optical lattice (experimental protocol shown in the top row). The bottom row shows the opening of a pseudogap in a normal-state attractive Hubbard system as the interactions are increased. \textbf{b} Spin-resolved bilayer readout. Bilayer physics and spin resolved readout in two dimensions can be realized using topologically protected transport in an optical superlattice. Prior to imaging, the planes of interest are transported away from each other, such that clean pictures of each plane (here with different spins in red/blue) can be obtained in a single experimental run. Finally, the spin resolved occupation of each site can be reconstructed. \textbf{c} Bilayer scheme to avoid parity projection during readout. The images below show the formation of a fermionic Mott insulator with one atom per site ($n=1$) in the wings and a central band insulator with $n=2$. The histogram on the right shows a clear separation for the single and double occupancy peaks. Panel a has been originally published in~\cite{brown2020}, panel b in~\cite{koepsell2020} and panel c in~\cite{hartke2020}.}
    \label{fig3}
\end{figure*}

Quantum gas microscopes continue to advance on the technical front to explore a broader range of physics. One recent advance has been using microscopes to probe spectral functions of Hubbard systems. Building on earlier work in unitary Fermi gases, the technique used emulates angle-resolved photoemission spectroscopy (ARPES) to probe single-particle excitations of the system~\cite{stewart2008, gaebler2010, sagi2015}. A radiofrequency photon flips one of the two interacting spin states of the Hubbard system into a third non-interacting state. The quasi-momentum of this final state is measured using bandmapping followed by a matter-wave focusing step that maps the atom's momenta to positions which can be measured with the microscope (see fig~\ref{fig3}a). Taking advantage of the high signal-to-noise of quantum gas microscopes, this technique has been used to measure the spectral functions of single layer attractive Hubbard systems~\cite{brown2020}. An alternative technique has been proposed that uses ``photoemission" by photon-assisted tunneling between occupied and empty partitions of the systems~\cite{bohrdt2018}. This eliminates the need for a non-interacting final state and allows the use of tailored optical potentials to measure local excitation properties. As cooling techniques advance, ARPES may be used to measure pseudogaps and changes in the topology of the Fermi surface in doped Mott insulators, complementing real space measurements.

Longer-range interacting many-body systems have been explored with quantum gas microscopes combined with optical  excitation. The short lattice spacing provides a relatively high but discrete density, allowing to achieve strong spin dependent interactions. Ising systems in different regimes have been realized by Rydberg excitation~\cite{schauss2012,schauss2015,zeiher2015,zeiher2016,zeiher2017a,guardado-sanchez2018}, precision spectroscopy in the lattice allowed the observation of huge Rydberg molecules held together exclusively by the van-der-Waals interaction one-by-one~\cite{hollerith2019} and collective atom-light interactions have been utilized to realize a sub-radiant ultra-light mirror~\cite{rui2020}.

Another recent advance is the introduction of bilayer systems (see fig.~\ref{fig3}b). On a technical level, these have been used to circumvent parity detection, which has allowed for example measuring the shell structure of bosonic Mott insulators and fluctuation thermometry in Fermi-Hubbard systems~\cite{preiss2015b,koepsell2020,hartke2020}. In the context of ARPES measurements, bilayers can provide a second empty layer for photoassisted tunneling. On the physics front, these systems offer the opportunity to emulate the physics of copper-oxide bilayers found in many high-temperature superconductors.

Finally, microscopy techniques are being extended to a wider range of species beyond the alkali metals. Quantum gas microscopes have been realized for ytterbium~\cite{miranda2015,yamamoto2016,miranda2017}, an atom with two-valence electrons, taking advantange of narrow-line optical molasses for cooling during imaging~\cite{yamamoto2016}. Ytterbium and alkaline-earth atoms are promising for the exploration of SU(N) quantum magnetism and topological physics~\cite{zhang2014,mancini2015,hofrichter2016,ozawa2018}. Microscopy apparatuses are under construction for highly magnetic atoms like erbium and bialkali ground-state molecules, both of which would be useful for the exploration of systems with long-range interactions. These microscopes may be used to probe exotic phases of matter like fractional Mott insulators and supersolids and for the exploration of novel dynamics like many-body localization with long-range interactions.  

\newpage

\bibliography{references}

\section*{Acknowledgements}
W.S.B. acknowledges funding from the National Science Foundation (grants no.  DMR-1607277, PHY-1912154), the David and Lucile Packard Foundation (grant no.  2016-65128), and the AFOSR Young Investigator  Research  Program  (grant  no. FA9550-16-1-0269). C.G. acknowledges funding from the European Union’s Horizon 2020 research and innovation program under grant agreement 817482 (PASQuanS), the European Research Council (ERC) 678580 (RyD-QMB), the Deutsche Forschungsgemeinschaft (SPP 1929 - GiRyd) and the Alfried Krupp von Bohlen und Halbach foundation.

\section*{Competing interests}
The authors declare no competing interests.

\end{document}